Trade-offs and synergies in managing coastal flood risk: A case study for New York City

Authors:

Robert L. Ceres, Department of Meteorology and Atmospheric Science, The Pennsylvania State University, University Park, Pennsylvania, USA, RobCeres@psu.edu

Chris E. Forest[1], Department of Meteorology and Atmospheric Science, Department of Geosciences, Earth and Environmental Systems Institute, The Pennsylvania State University, University Park, Pennsylvania, US, ceforest@psu.edu

Klaus Keller, Department of Geosciences, Earth and Environmental Systems Institute, The Pennsylvania State University, University Park, Pennsylvania, USA, klaus@psu.edu

## Abstract

Decisions on how to manage future flood risks are frequently informed by both sophisticated and computationally expensive models. This complexity often limits the representation of uncertainties and the consideration of strategies. Here, we use an intermediate complexity model framework that enables us to analyze a rich set of strategies, objectives, and uncertainties. We find that allowing for more combinations of risk mitigation strategies can expand the solution set, help explain synergies and trade-offs, and point to strategies that can improve outcomes.

## Introduction

Coastal communities are assessing the long-term risks from future storms and will choose among potentially expensive long-term risk mitigation strategies. These choices are often informed by state-of-the-art high-fidelity storm surge risk management modeling frameworks that evaluate options for at-risk regions such as New York City (NYC) (Aerts et al., 2013, 2014; Groves et al., 2016) and the Louisiana coastal region (J. R. Fischbach et al., 2012). The magnitude of proposed investments in these regions justifies extensive, site-specific research on both the future risks and the evaluation of risk-mitigation strategies. Current state-of-the-art high-fidelity modeling frameworks can pose considerable computational challenges in evaluating and optimizing a large number of strategies considering a wide range of future risks and to address a wide range of potentially conflicting objectives.

In NYC, an important goal of decisionmakers is to identify potential risk management strategies that cost-effectively satisfy the sometimes conflicting objectives of diverse stakeholder groups (New York City Special Initiative for Resilient Rebuilding, 2013; NYC Economic Development Corporation, 2014; Zhu & Lund, 2009). Previous risk mitigation proposals and risk mitigation strategy evaluations have broken important new ground, but the have assessed relatively few risk mitigation strategies and have not explicitly addressed the aggregate effectiveness of implementing combinations of strategies or quantitatively addressed many objectives (e.g., Mayors Office of Recovery & Resiliency, 2015; New York City Special Initiative for Resilient Rebuilding, 2013). The high computational costs associated with these frameworks and the limited computation budgets restricts the number of different strategies or strategy combinations that can be evaluated, the number of objectives that can be analyzed, and the extent of future risks that are considered. The relative sparsity of solutions combined with the limited number of objectives considered impose implicit a priori preferences to those objectives analyzed, thus potentially excluding a consideration of the preferences of key stakeholders. Previous research is often silent on the potential for improvements that can be achieved for many objectives by considering a more complex decision analysis. Additional factors to consider can include different levels of protection investment than those evaluated, additional strategies (such as flood insurance, implementation or restoration of natural or large-scale ocean barriers) and their associated policy levers, deep uncertainties driving the risks, or additional trade-offs associated with divergent stakeholders with potentially competing objectives.

---

[1] Corresponding author.

As an example, researchers in one study (Aerts et al., 2014) considered the NYC region and used a state-of-the-art storm surge risk mitigation framework consisting of a statistical/deterministic hurricane model using inputs from four climate models, two hydrodynamic models, and a wave height model. The Federal Emergency Management Agency's HAZUS damage models were then used to estimate damages. Using this method, the study evaluated four defensive strategies and one combined approach considering three objectives. Considered decision-levers included improvements to buildings to make them less vulnerable to storm damage, three storm surge barrier options, and a fifth strategy that combined building resistance with a barrier. The study estimated the life-cycle cost of the strategy and damage reduction to calculate a benefit-cost ratio (BCR) for each solution in three climate scenarios. The study found mixed results. Some of the strategies achieved BCRs ranging from the lowest of 0.13 under current climate conditions to 2.45 under a middle climate change scenario and a 4% discount rate. Other than BCR, additional stakeholder objectives and methods to improve these results were not explicitly addressed.

While a high BCR is desirable and often a necessary condition (USACE, 2018), stakeholders may value additional objectives (Gibbs, 2019) such as minimizing construction and maintenance costs, minimizing environmental impacts, or robustness of the strategy to uncertainties. The recently developed island City On a Wedge (iCOW) framework (Ceres et al., 2019) enables the analysis of high-dimensional objective spaces by using a simple storm surge model that evaluates a limited spatial domain. Because iCOW imposes relatively low computational costs, it can easily be coupled with multiple objective evolutionary algorithms. While iCOW is considerably less computationally expensive and easier to configure than high-fidelity modeling frameworks (Aerts et al., 2014; J. Fischbach et al., 2017), it incorporates many critical characteristics of US coastal communities. As a result, iCOW can be implemented at relatively low cost and is capable of evaluating and optimizing many combinations of risk management strategies, while considering many objectives representing the sometimes divergent priorities of many stakeholders. The iCOW framework is broadly adaptable to modeling many features typical of large coastal cities. Here, we modify the generic iCOW parameters to better reflect key features of Manhattan in terms of physical characteristics, demographics, and vulnerability. Next, we optimize three (3) combinations of storm surge risk mitigation strategies, using five (5) implementation levers, and develop a large set of Pareto optimal options considering six (6) objectives.

State-of-the-art storm surge modeling frameworks can provide useful high-fidelity representations of risk strategy effectiveness in a geospatially resolved region, but they impose high computational cost. The iCOW framework, in comparison, sacrifices a degree of realism and spatial resolution for computational efficiency. This trade-off makes iCOW useful for different classes of applications of interest to many community stakeholders. It offers the ability to explore many strategy combinations while considering many objectives, and it includes the ability to identify the full approximate Pareto front of optimal strategy combinations. This capability enables decisionmakers to consider more choices that may better satisfy the needs of diverse stakeholder groups.

### island City On a Wedge overview

The city simulated by the iCOW framework resembles the general physical and demographic profile of many major coastal cities, including Manhattan (Ceres et al., 2019). iCOW simulates a city along a waterfront and situated on a rising coast of constant slope. The city is initially uniformly dense, and buildings are uniformly tall (relative to the potentially largest storm surges). For this study, we use an iCOW model configured to resemble Manhattan. By modifying the original city-scape's defenses, we can analyze three (3) risk mitigation strategies that are implemented through five (5) strategy levers. For this proof-of-concept, we consider static strategies that do not adjust as new information may become available. Dynamic adaptive strategies have the potential to considerably improve outcomes (Garner and Keller, 2018, Haasnoot et al, 2013, 2019).

The first defensive strategy is withdrawal from low-lying at-risk areas. The second defensive strategy is to improve resistance by modifying buildings and infrastructures to reduce their vulnerability to flood damage. The third available strategy is to construct a dike (a.k.a. levee). Each strategy is defined by one or more levers that control the characteristics of the strategy. In this study, the three strategies considered are associated with five decision levers (see the XLRM framework used by Lempert et al., 2003). The withdrawal height lever is an elevation demarcation, below which buildings are relocated to higher elevation regions of the city. Resistance is associated with two levers, the resistance height and the resistance percent. The third strategy, building dikes, is associated with two levers, the dike height and the location of the dike base. Other mitigation strategies, such as insurance, enhancement of natural features that reduce the impact of storm surges, or harbor-scale surge barriers are possible, but for simplicity, are not included in this study.

Operation of the levers specifies a candidate city that is divided into distinct zones (Table SM1.1) with different values of densities and vulnerabilities to surge damage. For example, policymakers might adopt a strategy consisting of withdrawal from the lowest 0.2 m of the city, construction of a 5 m dike 1 m above the lowest city elevation, and implementation of building modifications to reduce flood damage by 50 % to a height of 5 m for the buildings located between the withdrawal area and the dike. In this case, the candidate city will consist of zones zero, one, three, and four from Table SM1.1.

We evaluate candidate city strategies against possible future storm surges using objectives that different stakeholder communities might have. For this study, we consider six objectives: (1) minimize total investment cost, (2) minimize average annual damage (over a 50 year period), (3) maximize return on investment (ROI), (4) maximize total monetary net benefit of investment, (5) maximize the frequency of a positive total monetary net benefit, and (6) minimize the annual frequency of large threshold events.

To calculate these objectives, we generate a series of storm surges that represent the annual highest water levels at The Battery tide gauge located adjacent to Battery Park in lower Manhattan. We use the Generalized Extreme Value distribution (GEV) and adjust its parameters such that the 100-year storm surge level increases by 1 m per century (Ceres et al., 2019). Next, we use a multiple objective evolutionary algorithm (MOEA) (Hadka & Reed, 2012, 2013) to generate an initial random population of candidate strategies, each defined by their associated lever settings and evaluate the objectives. The MOEA then recombines the traits of successful population members to evolve the population of candidate strategies towards the Pareto-optimal surface spanned by the objectives. We use data visualization tools to illustrate the potential trade-offs and synergies.

## Results and Discussion

The iCOW framework maps out the approximate Pareto front. We discuss a sample of strategies located at this Pareto front.  We identify a zero investment (`do nothing') strategy, that result in $389 million in damages per year. The most expensive ('all in') strategy costs $6.5 trillion to implement and is associated with an average estimated annual damages of $6.6 million/yr. This all in strategy is dike-free and includes a withdrawal from the waterfront to 0.33 m and unprotected buildings made 99.5% resistant to damage to a height of 13 m above the withdrawal height. A "lowest damage" strategy combines a withdrawal from the waterfront to 0.37 m and unprotected buildings made 99.5% resistant to damage to a height of 13.9 m above the withdrawal height. There is a 10 m dike located 5.9 m above the withdrawal height. This strategy cost $4.3 trillion but results in estimated damages of just $0.3 million/yr.

Optimization of the fully parameterized iCOW Manhattan model results in a diverse Pareto dominant solution set consisting of 18,756 members. We show the Pareto front for all solutions requiring less than $25 billion investments (Figs. 1-3). In comparing the results with those generated using two objectives (Ceres et al., 2019), we find that the inclusion of additional objectives results in a more diverse and larger solution set.

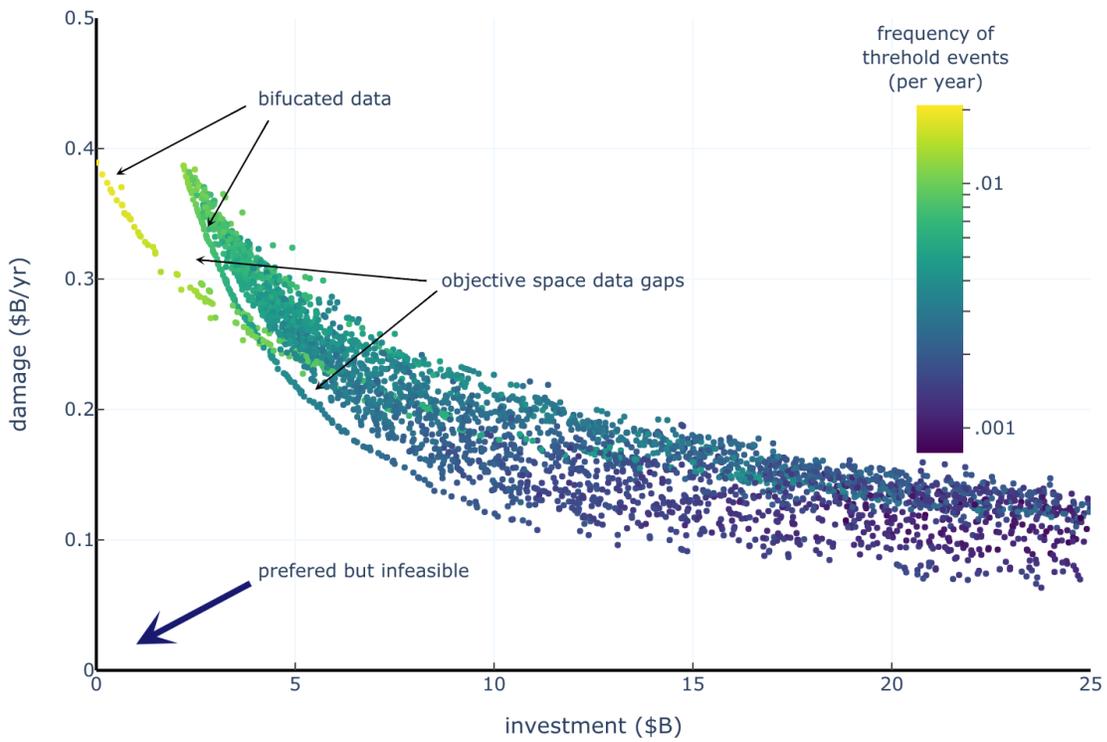

*Figure 1 Two dimensional data visualization of Manhattan iCOW emulation showing damage (y axis) vs. investment (x axis) for the Pareto dominant solution with investment cost less than $25 billion. The color scale shows Pareto dominant values for the annual frequency of threshold events. Average total monetary net benefit, frequency of positive total monetary net benefit, and benefit to cost ratio are not shown.*

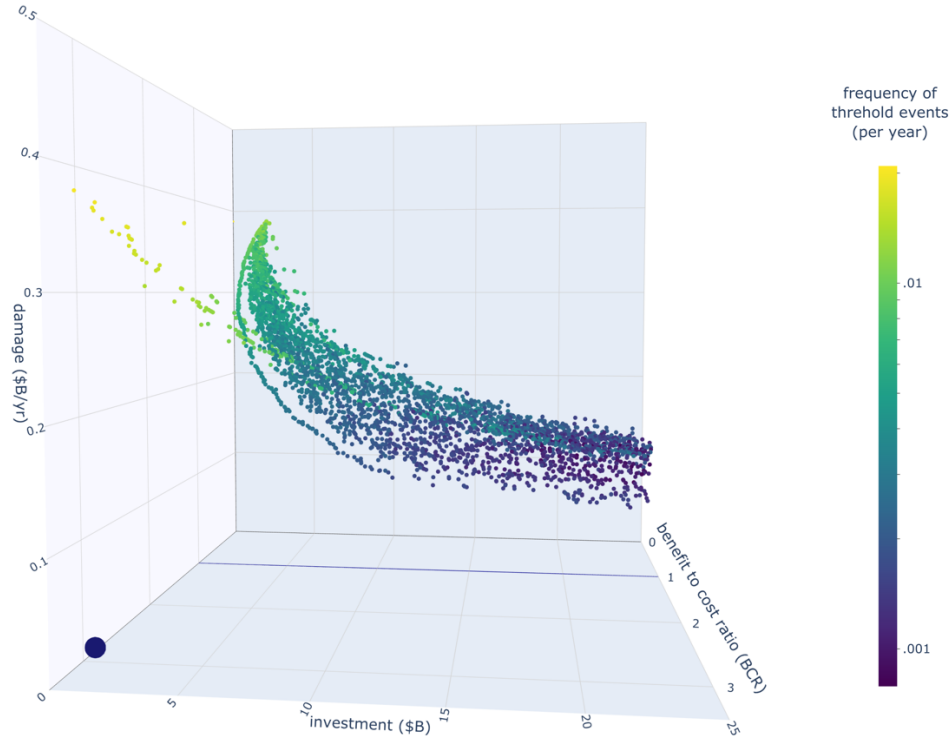

*Figure 2 Three-dimensional data visualization of Manhattan iCOW emulation showing damage (z axis) vs. investment (x axis) vs. benefit to cost ratio (y axis) for the Pareto dominant solution with investment cost less than $25 billion. Color scale shows Pareto dominant values for the annual frequency of threshold events. The most preferred solutions (near zero cost and damage with a low frequency of threshold events and high benefit to cost ratio are located in the lower left forward corner of the plot marked with the large dark blue circle.*

**The Need for More Complex Data Visualizations**

Many iCOW two axis objective plots clearly reveal the Pareto dominant front for the two axis objectives as a well-defined edge closest to each axis' objective goal (e.g. zero damage and no investment cost in Fig. 1). Points fanning behind this front are associated with other objectives, but the two-dimensional plot fails to illustrate the trade-offs associated with these other objectives. Encoding information about these objectives via adding a third axis, a color scale, or a point-size scale may help or hinder policymaker understanding, as can additional 2D plots showing different objective trade-offs. Figs. 2 and 3, for example, provide a far better representation of the bifurcation and gaps in the data seen in Figs. 1 and 3. By providing a visual representation of trade-offs, the additional information may help, but the associated visual complexity may create further barriers to understanding what is already a complicated problem. Using a three-axis plot, we allow for the display of an additional objective (as in Fig. 2 and supplemental Fig. SM1.1), at the cost of potentially obfuscating the relationships between objectives due to the projection of a complicated three-dimensional shape onto a two-dimensional surface. An effective approach that overcomes many of these limitations may be to allow for interactive data visualization techniques where the user can adjust the view to their preferences. (See supplemental material for additional examples and discussion.)

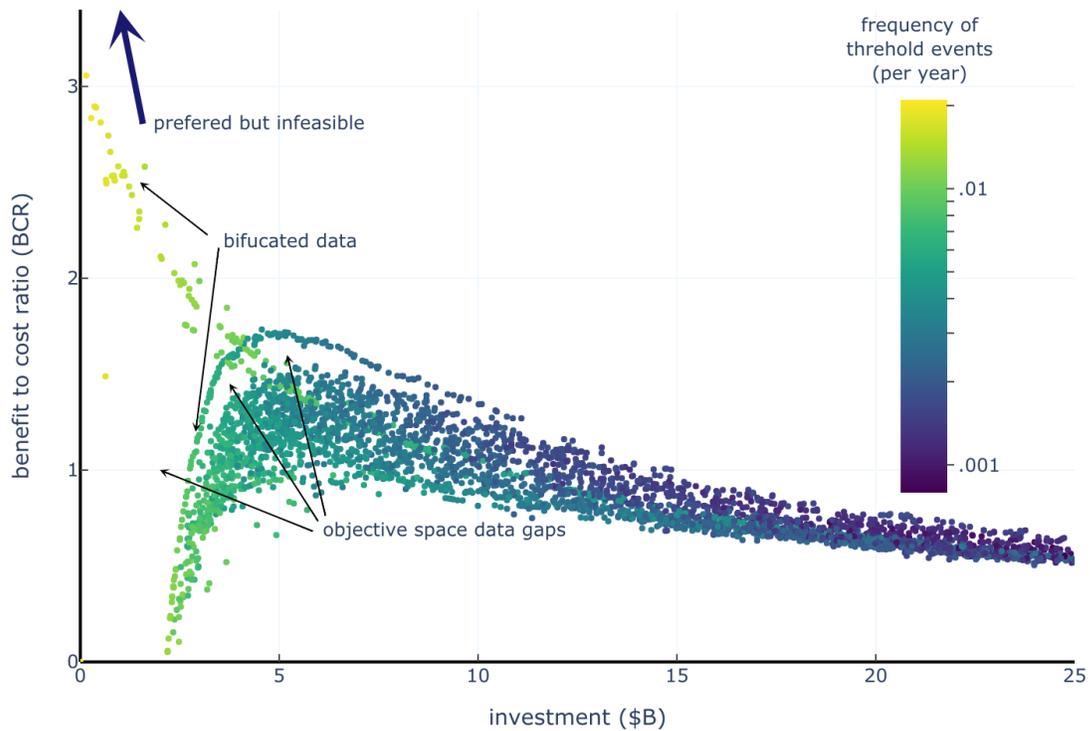

*Figure 3 Two-dimensional data visualization of Manhattan iCOW emulation showing benefit to cost ratio (BCR) (y axis) vs. investment (x axis) for the Pareto dominant solution with investment cost less than $25 billion.*

Alternative data visualization approaches, such as parallel axis and matrix plots, can help highlight the complicated inter-dependencies and trade-offs between the levers and objectives. As one example, a parallel axis plot (see supplemental material) can show the richness of solutions in a Pareto front. However, this display does not always allow for clear identification of Pareto fronts (see the supplemental material for a discussion on the causes for the visual features as well as different plots showing the relationships between different objective combinations).

The Pareto dominant population assumes no a priori preference for objectives and spans a wide range of objectives and lever settings. Not all stakeholders have the ability to influence all of the levers, and resources to implement levers may be limited. For example, a city government may have a fixed range of investments available for capital investment in structural solutions such as constructing dikes or levees. They also may have some control over other levers such as the ability to impose reasonable zoning restrictions that improve structural resistance to damage. And, they may have no control over the withdrawal of private infrastructure to higher locations in the city. Other stakeholders may have strong preferences that city governments may wish to consider. For instance, the waterfront business community may not prefer dikes or levees located at the waterfront that restrict access to their business locations. The data visualization can provide controls that allow stakeholders with divergent objectives to mutually explore and clearly illustrate the trade-offs inherent in the Pareto dominant solution space.

**Implications**

Our analysis drastically increases the number of different strategies compared to previous studies (Aerts et al., 2014; J. Fischbach et al., 2017). These previous studies are often silent on whether the particular combination of strategies evaluated can be improved (in terms of one or more objectives) by changing some or all of the implemented strategy levers. For example, one study (Aerts et al., 2014) examined five strategy combinations for New York City costing $10.4 - $23.8 Billion with a range BCRs ranging from 0.13 under current climate conditions to 2.45 under a moderate climate change scenario and a 4 % discount rate. Moreover, this study is silent on the question whether each of the individual strategy combinations studied could be improved through changes to investment levels or changes in implementation details.

iCOW can act as a scout to identify promising new approaches that can then be checked with more refined models. Our study suggests that there may be lower-cost resistance-only based strategies (Fig. SM19) that have very high benefit to cost ratios and higher frequency of positive net benefit that were not identified or evaluated in previous studies using more complex model (such as Aerts et al., 2013).  These new strategies may be preferred by stakeholders. This added insight from the iCOW results, however, comes at the cost of providing less and lower fidelity information regarding any particular solution.

The iCOW framework estimate of expected annual storm surge damage over the next 50-years for the undefended model ($389 million/yr) is substantially larger than the $174 million/yr estimate for NYC (the entire city) from previous research (Aerts et al., 2013). Many factors may contribute to this difference. For example, one potential source for the larger damages found in this study is that the GEV location and scale parameters are increasing such that the 100-year storm surge is increasing at one meter per century. Alternatively, we can express this imposed change in terms of shorter return periods for major surges. For the observed 2.8 m surge from Hurricane Sandy, for instance, the estimated return period decreases from 130 years to 60 years for the 50$^{th}$ year in the simulations considered here. We have repeated the analysis for the case where GEV parameters are held constant at current estimates and results are qualitatively similar (see supplemental material). Another source for larger damages may result from iCOW's assumptions regarding seawall height (which precludes most storms from causing any damage), and iCOWs assumed linear relationship between city volume flooded and damage. The net effect of these assumptions is that damage rises very rapidly for surges larger than Superstorm Sandy's 2.8 m (above MHHW) surge (see supplemental material section SM1). Two-dimensional data visualization of the iCOW Framework's data output may obscure important characteristics of the solution space, thus not supporting decisionmakers and stakeholders' need for clear and decision-relevant information.

Taking advantage of the iCOW framework's capabilities to support actual decisionmaking for NYC (or any city) will require improvements to the overall fidelity of the iCOW framework beyond this proof-of-concept study. City policymakers and stakeholders can draw upon the city's considerable collective experience and local expertise to identify factors that are specific to local conditions. This additional knowledge may justify changes to the iCOW framework, such as adjustments to cost parameters, cost algorithms or damage functions for dikes, resistance, and withdrawal. Stakeholders may wish to consider additional risk mitigation strategies not considered in this study, such as expanded insurance coverage, enhancement of natural barriers, or city resiliency improvements. For example, the current damage functions are calibrated to result in zero damage for surges below the seawall height, and to generate approximate `Sandy-scale' damages from `Sandy-scale' surges. Results from other model frameworks (Aerts et al., 2013; Lin et al., 2012) can be used to explore the iCOW framework's emulation performance and to identify, quantify, investigate, and potentially remedy sources of deviation. In addition, local expert knowledge often suggest additional decision levers such as installation of pumps, or a strategy to not rebuild after damage). Similarly, they may also have additional site-specific objectives (such as defense of specific critical infrastructure).

## Conclusions

The iCOW framework can be useful to decisionmakers in multiple ways. First, it can examine and optimize many "real-world" strategy options over a wide range of uncertainties to generate a diverse solution set that considers many objectives. This capability can identify additional options for state-of-the-art solutions where multiple objectives have not been considered. Second, many coastal communities lack the resources required to employ difficult and expensive to implement state-of-the-art storm surge risk modeling frameworks. In these cases, the iCOW framework can be employed relatively inexpensively and quickly to support coastal community decisionmaking.

## Methods

To assess the iCOW framework's ability to adequately model Manhattan's storm surge risk, we evaluate the fully parameterized iCOW model of an undefended city against a range of storm surges, up to and exceeding surge levels experienced during Superstorm Sandy. We also use the full iCOW framework to assess the overall performance of the unprotected city against the aggregate exogenous sequences of storm surges and compare the performance of the unprotected city to established risk estimates currently being used by NYC and estimated by other research.

To test the framework's ability to inform stakeholders evaluating storm surge risk mitigation strategy options, we evaluate and optimize the six objectives using the five strategy levers against the exogenous storm surges. We conduct this optimization using 100 cores in parallel, and run the optimization for 12 hours, resulting in resultant population of more than 18,000 members.

We project the results using two and three-dimensional plots and parallel axis plots. However, different stakeholders with differing objectives and abilities to affect decisions may have different ideas on how to display data to best visualize trade-offs among their priorities. Moreover, using an interactive display often improves comprehension of complex and high dimensional data. Therefore, we will also provide an online visualization tool to interactively explore the data sets developed for this study at datacommons.psu.edu after publication of this article.

## Code and data availability

All iCOW software code and the final dataset used to create the figures in this article are available at (https://github.com/rceres/ICOW_Manhattan). This is a research study and the results are not to be used to inform actual decision-making. All results, data, and tools are distributed under the GNU non-commercial license 3.0 (https://www.gnu.org/licenses/lgpl-3.0.html) (see especially the disclaimer of warranty and limitation of liability).

## Acknowledgments

We thank D. Hadka for outstanding technical support with the Rhodium multi-objective tool kit and the Borg MOEA. We thank B. Lee, R. Lempert, and J. Lawrence for helpful discussions. This research was partially supported by the National Science Foundation (NSF) through the Network for Sustainable Climate Risk Management (SCRiM) under NSF cooperative agreement GEO-1240507 and the Penn State Center for Climate Risk Management. Any opinions, findings, and conclusions or recommendations expressed in this material are those of the authors and do not necessarily reflect the views of the funding entities.

## Author contributions

All authors contributed to the island City on a Wedge (iCOW) framework conceptual design. RC developed the iCOW framework, detailed model designs, and experiment designs, wrote all software code, and integrated iCOW software with the BORG MOEA algorithm and J3 data visualization toolkit. RC executed all software and conducted all analyses. All authors developed the experimental plan. CF and KK provided guidance for the project. RC wrote the article and all authors contributed to editing.

## Conflict of interest

The authors are not aware of financial or personal relationships that would pose a conflict of interest.

# References

Aerts, J. C. J. H., Botzen, W. J. W., Emanuel, K., Lin, N., de Moel, H., & Michel-Kerjan, E. O. (2014). Evaluating Flood Resilience Strategies for Coastal Megacities. *Science*, *344*(6183), 473–475. https://doi.org/10.1126/science.1248222

Aerts, J. C. J. H., Lin, N., Botzen, W., Emanuel, K. A., & de Moel, H. (2013). Low-Probability Flood Risk Modeling for New York City: Low-Probability Flood Risk Modeling for New York City. *Risk Analysis*, *33*(5), 772–788. https://doi.org/10.1111/risa.12008

Ceres, R. L., Forest, C. E., & Keller, K. (2017). Understanding the detectability of potential changes to the 100-year peak storm surge. *Climatic Change*, *145*(1–2), 221–235. https://doi.org/10.1007/s10584-017-2075-0

Ceres, R. L., Forest, C. E., & Keller, K. (2019). Optimization of multiple storm surge risk mitigation strategies for an island City On a Wedge. *Environmental Modelling & Software*, *119*, 341–353. https://doi.org/10.1016/j.envsoft.2019.06.011

Coles, S. (2001). *An Introduction to Statistical Modeling of Extreme Values. Springer Series in Statistics.* Springer Veriag.

de Blasio, B., & Joeseph F. Bruno. (2014). *New York City Hazard Mitigation Plan (2014)*. Hazard Mitigation Unit, New York City Office of Emergency Management. http://www.nyc.gov/html/oem/downloads/pdf/hazard_mitigation/plan_update_2014/1_introduction_final.pdf

Fischbach, J., Johnson, D., & Molina-Perez, E. (2017). *Reducing Coastal Flood Risk with a Lake Pontchartrain Barrier*. RAND Corporation. https://doi.org/10.7249/RR1988

Fischbach, J. R., Louisiana, & Rand Gulf States Policy Institute (Eds.). (2012). *Coastal Louisiana risk assessment model: Technical description and 2012 coastal master plan analysis results*. Rand Corp.

Garner, G. G., & Keller, K. (2018). Using direct policy search to identify robust strategies in adapting to uncertain sea-level rise and storm surge. Environmental Modelling & Software, 107, 96–104. https://doi.org/10.1016/j.envsoft.2018.05.006.

Gibbs, K. (2019). *EP_ 1105-2-57-2 Stakeholder engagement, collaboration, and coordination*. US Army Corp of Engineers. www.publications.usace.army.mil/Portals/76/Users/182/86/2486/EP_%201105-2-57.pdf?ver=2019-04-03-150516-977

Groves, D., Kuhn, K., Fischbach, J., Johnson, D., & Syme, J. (2016). *Analysis to Support Louisiana's Flood Risk and Resilience Program and Application to the National Disaster Resilience Competition*. RAND Corporation. https://doi.org/10.7249/RR1449

Hadka, D., & Reed, P. (2012). Diagnostic Assessment of Search Controls and Failure Modes in Many-Objective Evolutionary Optimization. *Evolutionary Computation*, *20*(3), 423–452. https://doi.org/10.1162/EVCO_a_00053

Hadka, D., & Reed, P. (2013). Borg: An Auto-Adaptive Many-Objective Evolutionary Computing Framework. *Evolutionary Computation*, *21*(2), 231–259. https://doi.org/10.1162/EVCO_a_00075

Haasnoot, M., Kwakkel, J. H., Walker, W. E., & ter Maat, J. (2013). Dynamic adaptive policy pathways: A method for crafting robust decisions for a deeply uncertain world. Global

Environmental Change: Human and Policy Dimensions, 23(2), 485–498. https://doi.org/10.1016/j.gloenvcha.2012.12.006

Haasnoot, M., Brown, S., Scussolini, P., Jimenez, J., Vafeidis, A. T., & Nicholls, R. (2019). Generic adaptation pathways for coastal archetypes under uncertain sea-level rise. Environmental Research Communications. https://doi.org/10.1088/2515-7620/ab1871

Lempert, R.J., Popper, S., Bankes, S., 2003. Shaping the Next One Hundred Years:New Methods for Quantitative, Long Term Policy Analysis. RAND Corporation, Santa Monica,CA

Lin, N., Emanuel, K., Oppenheimer, M., & Vanmarcke, E. (2012). Physically based assessment of hurricane surge threat under climate change. *Nature Climate Change*, *2*(6), 462–467. https://doi.org/10.1038/nclimate1389

Mayors Office of Recovery & Resiliency. (2015). *Nyc_ndrc_phase2_english.pdf*. New York City.

New York City Special Initiative for Resilient Rebuilding. (2013). *PlaNYC: A Stronger, More Resilient New York* [Special Report]. http://www.nyc.gov/html/sirr/html/report/report.shtml.

NOAA. (n.d.). *Water Levels—NOAA Tides & Currents*. NEW YORK (The Battery), NY StationId: 8518750. Retrieved October 9, 2014, from http://tidesandcurrents.noaa.gov/waterlevels.html?id=8518750&units=metric&bdate=20121029&edate=20121030&timezone=GMT&datum=MHHW&interval=6&action=

NYC Economic Development Corporation. (2014). *Southern Manhattan Coastal Protection Study: Evaluating the Feasibility of an MPL*.

Reed, A. J., Mann, M. E., Emanuel, K. A., Lin, N., Horton, B. P., Kemp, A. C., & Donnelly, J. P. (2015). Increased threat of tropical cyclones and coastal flooding to New York City during the anthropogenic era. *Proceedings of the National Academy of Sciences*, *112*(41), 12610–12615. https://doi.org/10.1073/pnas.1513127112

Scileppi, E., & Donnelly, J. P. (2007). Sedimentary evidence of hurricane strikes in western Long Island, New York. *Geochemistry, Geophysics, Geosystems*, *8*(6), n/a-n/a. https://doi.org/10.1029/2006GC001463

Sweet, W., Zervas, C., Gill, S., & Park, J. (2013). Hurricane Sandy innundation probabilities today and tomorrow. *Bulletin of the American Meteorological Society*, *94*(9), S17–S20.

US Census Bureau. (2018). *U.S. Census Bureau QuickFacts: New York County (Manhattan Borough), New York*. https://www.census.gov/quickfacts/fact/table/newyorkcountymanhattanboroughnewyork/BZA210216

USACE. (2018). *ASCE Federal Project BCR and Scoring Information Paper*. US Army Corp of Engineers. https://www.asce.org/uploadedFiles/News_Articles/asce-bcr-paper-2018.pdf

*Water Levels—NOAA Tides & Currents*. (n.d.). Retrieved September 5, 2018, from https://tidesandcurrents.noaa.gov/waterlevels.html?id=8518750&units=metric&bdate=19600912&edate=19600912&timezone=GMT&datum=MHHW&interval=h&action=

Zhu, T., & Lund, J. R. (2009). Up or Out?—Economic-Engineering Theory of Flood Levee Height and Setback. *Journal of Water Resources Planning and Management*, *135*(2), 90–95.

Zitzler, E., Thiele, L., Laumanns, M., Fonseca, C. M., & da Fonseca, V. G. (2003). Performance assessment of multiobjective optimizers: An analysis and review. *IEEE Transactions on Evolutionary Computation*, *7*(2), 117–132. https://doi.org/10.1109/TEVC.2003.810758

Supplementary Material

SM1 iCOW Framework technical description

For a modeled city, the iCOW Framework attempts to optimize candidate cities considering many potential objectives and using many combinations of defensive strategies. The goal of this optimization is to better inform stakeholders and decisionmakers about the trade-offs associated with different combinations of defensive strategies without any a-priori preference for particular objectives. The iCOW framework accomplished this optimization through the interactions of three modules (exogenous inputs, the City Model, and multiple objective evolutionary algorithms (MOEA)), and a data visualization toolkit (see Fig. SM1) as described briefly in sections SM1.1 and 3.2 below.

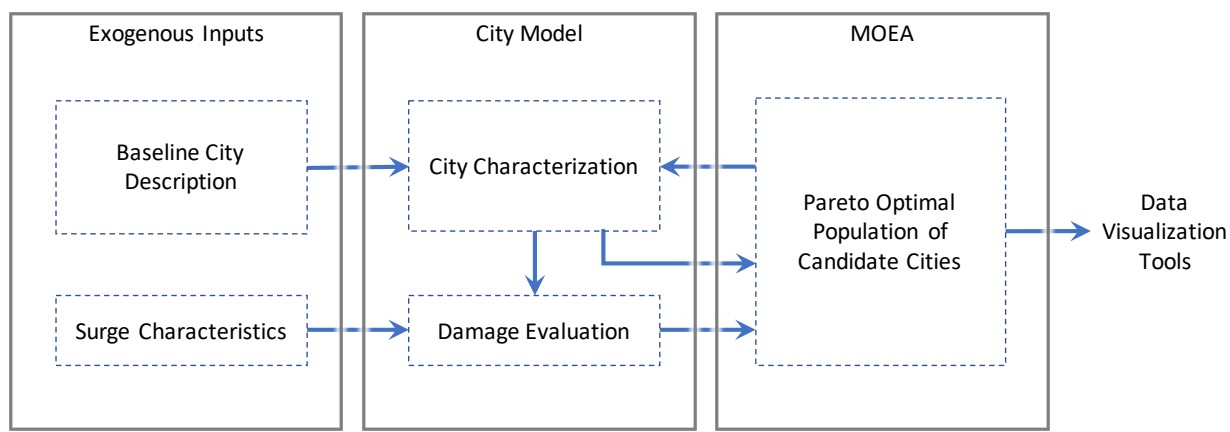

*Figure SM1. The island City On a Wedge (iCOW) framework consists of three modules: exogenous inputs, the City Model, and Multiple Objective Evolutionary Algorithms (MOEA).*

We use the iCOW framework to evaluate the effectiveness of three defensive strategies aimed at reducing storm surge impact, withdrawal, resistance, and dikes(Ceres et al., 2019). Other mitigation strategies, such as insurance, enhancement of natural features that reduce the impact of storm surges, or harbor-scale surge barriers are possible, but for simplicity, are not included in this study.

Operation of the five strategy levers specifies a candidate city that is divided into distinct zones (listed in Table SM1) with different values of densities and vulnerabilities to surge damage. For example, policymakers might adopt a strategy consisting of withdrawal from the lowest 0.2m of the city, construction of a 5m dike or levee 1m above the lowest city elevation, and implementation of building modifications to reduce flood damage by 50% to a height of 5m for the buildings located between the withdrawn area and the levee. The candidate city will then consist of zones zero, one, three, and four.

Table SM1. Relationship between zones, strategies, and levers

| Zone | Name | Description | levers |
|---|---|---|---|
| 0 | Withdrawal zone | Area above the seawall where infrastructure has been moved to higher ground. | **W**ithdrawal height |
| 1 | Resistant zone | Area above the withdrawal zone where the lower portions of buildings are made more resistant to damage. | **R**esistance height resistance **P**ercent |
| 2 | Unprotected zone | Area above the Resistant zone but below the dike base that is unprotected | n/a |
| 3 | Dike zone | Area behind and protected by the dike. | **Di**ke height dike **B**ase height |
| 4 | City heights | Area above the dike top that is unprotected. | n/a |

The iCOW framework evaluates the effectiveness of defensive strategies, i.e. the city zones, using metrics corresponding to objectives differing among stakeholder communities. For this study, we use six output metrics: total investment cost, average annual damage (over a 50-year period), benefit to cost ratio (BCR), total monetary net benefit of investment, frequency of a positive total monetary net benefit, and the annual frequency of large threshold events.

Given these objectives, the framework seeks to identify a population of candidate cities (described by their lever lettings and associated objective metrics) that approach the best achievable combinations of objectives without preference to the objectives. A Pareto optimal solution exists when it is not possible to improve on the performance of one objective without degrading the performance of one or more other objectives. The full set of optimal solutions describes a Pareto optimal solution space or front.

SM1.2. Exogenous inputs

The city simulated by the iCOW framework resembles the general physical and demographic profile of many major coastal cities, including Manhattan(Ceres et al., 2019). We simulate a city along a waterfront and situated on a rising coast of constant slope. The city is initially uniformly dense, and buildings are uniformly tall (relative to the potentially largest storm surges). The first exogenous input is the set of parameters describing the baseline city. Selection of parameters to simulate Manhattan is discussed in section SM1.5 and parameter values are listed in Table SM2.

The second exogenous input is the sequence of storm surges used to calculate storm surge damages. For this study, we generate annual highest storm surges representative of The Battery tide gauge from the Generalized Extreme Value distribution (GEV) as described in (Ceres et al., 2017). We adjust the GEV parameters such that the 100-year storm surge level increases by 1m per century for consistency with (Ceres et al., 2017, 2019).

SM1.3 City model

The City Model has two major components evaluated in sequence. The first component takes the exogenous factors (described above) and the five strategy levers corresponding to three defensive strategies (Table SM2) and characterizes the candidate city (defined as the original modeled city as modified by the lever settings) in terms of the value distribution, zones, and associated damage functions. The costs to implement the strategy lever inputs is a City Model output metric. The second component takes the city's value distribution, zones, and damage functions as inputs from the MOEA module (see section SM1.3 and evaluates the city's response to the exogenous set of storm surge sequences, which generate the remaining module output metrics. For additional technical details, see (Ceres et al., 2019).

The defensive strategies for a candidate city are established at time zero, and hence, the first component to characterize the city is evaluated only once. The City Model evaluates each candidate city against every storm surge for all states of the world considered. We use six objectives, the total cost of implementing the combined defensive strategies, average damage per year, benefit to cost ratio (BCR), total monetary net benefit, the frequency of achieving a positive total monetary net benefit, and the annual frequency of storms exceeding a damage threshold. The objectives are evaluated over a 50-year time span. Total monetary net benefit represents the difference between the amount of damage reduction and investment cost. BCR is damage reduction divided by investment cost. (see additional details in the supplemental material).

The City Model output metrics are calculated using 5,000 50-year time periods, (aka states of the world) for each candidate city. Output metrics for candidate cities are returned as the input metrics for the MOEA discussed in section SM1.3 below. For this study, we also output the additional information metrics of the cost to implement each strategy. Many other information metrics are available or could be easily implemented for tracking or use as objectives for optimization.

We select iCOW model parameter values based on three goals in descending order of priority. The first goal is to plausibly match the physical characteristics of Manhattan. The second goal is to match City Model outputs to predicted costs or estimated responses to the Borough of Manhattan as stated by NYC. The third goal is to ensure that iCOW framework results are comparable to results from other research initiatives as well as estimates of the cost and effectiveness of ongoing efforts to make Manhattan more resistant to storm surges.

The Borough of Manhattan is situated on Manhattan Island. According to the US Census bureau, the borough is approximately 21.6 km long, is approximately 3.7 km wide at its widest point, has an area of 59.2 km$^2$ (US Census Bureau, 2018) and is surrounded by a seawall or natural elevation barriers.

iCOW models this geometry as a rectangular wedge as depicted in Fig. SM1. To make for a good conceptual fit, we consider the east and west sides of the borough as a single coastline and adjust the dimension to achieve a reasonable representation of the irregularly shaped Manhattan with the rectangular shape of the City Model. We therefore select a city width of 40 km, and a city width of 1.5 km, resulting in a total area of 60 km$^2$. We use a representative seawall height of 2m. For further details, see Supplemental section SM1.4 and Supplemental table SM2.

Our projection of damage resulting from Superstorm Sandy, $6.4 billion, is roughly consistent with estimates of actual damage to Manhattan ($7.0 billion)(de Blasio & Joeseph F. Bruno, 2014). Average annual damage to the undefended iCOW city over the full 5,000 sets of 50-year storm surge sequences, ($374 million/yr) however, are substantially higher than those estimated by other studies(Aerts et al., 2014; Reed et al., 2015). We also estimate the risk of future threshold events (resulting in damages of \$4 billion or more).

SM1.4 Multiple Objective Evolutionary Algorithms (MOEA)

Conceptually, the MOEA module generates an initial random population of candidate cities, uses the City Model to evaluate each population member, then recombines the traits of successful population members to evolve the population of candidate cities towards improved performance. See (Zitzler et al., 2003) for more details on how MOEAs function. The module evaluates the success of each candidate city in terms of the six objectives described above. We selected the BORG MOEA based on its excellent performance in optimizing problems with discrete choices, exploring and generating diverse solutions in problems with multi-modal solution regions, and its computational efficiency(Hadka & Reed, 2013). Note that the goal of using MOEAs is to generate a diverse population that, without preference for particular objectives, converges to the full trade-off surface for all the objectives. We use the term Pareto dominant solution to refer to a population where the member's objective performance is as good or better than the preceding populations, and thus are converging towards the Pareto optimal surface.

SM2 Manhattan and development of iCOW parameters

The Battery Park seawall is approximately 1.2 m above the mean high high water (MHHW) mark of the NOAA tide gauge located adjacent to the park(NYC Economic Development Corporation, 2014). The authors are not aware of other published data on the average height of the seawall surrounding Manhattan, but visual observation indicates that it is usually somewhat higher. In the iCOW model no damage occurs when storm surges are below the seawall height. The storm surge from Superstorm Sandy reached 2.8 m above MHHW at The Battery (see supplemental Fig. SM6), and caused extensive flooding. Whereas the storm surge from Hurricane Donna in 1960 reached just over 1.6 m (see supplemental Fig. SM7), but did not cause enough flooding and damage to be measured and recorded in NYC reports. Similarly, the winter storm of November 1960, nicknamed the Great Appalachian Storm, reached a peak storm surge of 2.3 m, but the storm tide only reached 1.2 m above MHHW (see supplemental Fig. SM8 ). This storm did cause some flooding in Manhattan, but did not result in extensively reported damage. We therefore establish a uniform iCOW seawall height of 2.0 m as a reasonable representative seawall height. See supplemental material for additional discussion on observed storm surges and their associated storm tides.

The overall topography of Manhattan resembles a ridge oriented parallel to the long (north south) axis of the borough. A plaque situated in Bennett Park marks Manhattan's highest point and lists the elevation as 265 ft. The peak elevation of the ridge at other locations is considerably lower, and as elevation from the waterline increases, the rectangular shape of the City Model less faithfully models the actual borough terrain. However, above the maximum surge heights, City Model objectives, such as damage functions, are not affected by this mismatch. We therefore derive a representative iCOW city height based on NYC's reported inundation extent from Superstorm Sandy of $P_i$ = 11% of buildings damaged (for all five boroughs). From this percentage, and the observed Superstorm Sandy surge height of $S_{Sandy}$ = 2.8 m, we derive a representative iCOW peak elevation of 17 m,

$$E_{Representative\ max} = \frac{S_{Sandy}}{P_{area\ inundated}} \cong 17\ m.$$

Not all buildings within a flooded area are necessarily damaged. In Manhattan, Superstorm Sandy's storm surge damaged 39% of buildings within the inundation footprint(de Blasio & Joeseph F. Bruno, 2014). We therefore establish a parameter $F_{damaged}$ = 0.39 to represent this fraction.

The remaining city parameters of the uniform city building height, average structure vulnerability, and total city value are interrelated. We assume a uniform iCOW building height, $h_{building}$ = 30 m. This height is higher than the maximum storm surges, but considerably shorter than the tallest buildings in Manhattan.

The authors know of no definitive estimates of Manhattan's total asset value. Therefore, we develop a representative overall value for Manhattan, $V_{city}$, based on iCOW dimensions, and NYC's estimates of inundation extent and reported damage associated with Superstorm Sandy in accordance with relationship,

$$\frac{v_{damage\ from\ Sandy}}{v_{city}} = \frac{P_{volume\ innundated}}{V_{city}}, \tag{2}$$

where,

$$P_{volume\ innundated} = \frac{(surge_{Sandy} - h_{seawall})^2}{2}, \tag{3}$$

and,

$$V_{city} = h_{buildings} * city\ width * E_{representitive\ max}. \tag{4}$$

The resultant representative total city value is $1.5 trillion.

the above parameters are listed in Table SM1. The remaining parameters relating to the cost and damage functions for the iCOW city zones are as described in (Ceres et al., 2019). Data within the scientific literature on these parameters is sparse, but more realistic city specific parameters could be developed by local policy makers. For instance, the construction costs for a levee or dike will vary regionally based on labor rates, site accessibility, local environmental concerns, material costs, degree of exposure to open seas, or ground and soil conditions. Local policymakers and stakeholders will typically have better estimates of these factors which should be incorporated to improve the overall fidelity of the iCOW framework.

Table SM2. Manhattan iCOW parameters

| Parameter | Value | Units |
| --- | --- | --- |
| Building height | 30 | m |
| City elevation | 17 | m |
| City depth | 2 | km |
| City length | 43 | km |
| City value | 1.3 | $ trillion |
| Damage factor | 0.39 | none |

The BORG MOEA uses the concepts of epsilon-dominance and an epsilon-box strategy to prevent deterioration of the Pareto dominant candidate population away from local Pareto optimal solutions and to provide diversity in the final Pareto dominant population (Hadka & Reed, 2013). Improper selection of epsilon parameters can affect performance of the BORG MOEA algorithm by increasing the time required to archive convergence or by preventing identification of locally optimal combinations of lever settings, thus decreasing population diversity (Hadka & Reed, 2012). From the pragmatic stakeholder or policymaker perspective, epsilon parameters can be thought of as the approximate maximum difference in an objectives value that would otherwise be of little or no consequence. Additionally, poor selection of epsilon parameters and the subsequent distribution associated with the converged Pareto dominant solution cities can obscure potentially important trade-offs between objectives. Table SM3 lists epsilon parameters used in the BORG MOEA for this study.

Table SM3 iCOW MOEA parameters

| Epsilon parameter | Value | Units |
| --- | --- | --- |
| investment | 100 | $ million |
| damage | 100 | $ million |
| BCR | 0.02 | none |
| total monetary net benefit | 10 | $ million |
| freq. threshold events | 0.0002 | none |
| freq. positive total monetary net benefit | 0.01 | none |

SM3 Discussion of storm surge

Individual return period estimates for a Superstorm Sandy's storm tide at the battery range typically exceed 500 years and are associated with large uncertainty. For examples, Fig. SM2 shows The Battery tide gauge modeled return levels from synthetically storms generated in (Lin et al., 2012) and later used in (Aerts et al., 2013). Fig. SM3 shows model fit to observations at The Battery from (Sweet et al., 2013). Super Storm Sandy's surge observation visually stands out as a potential outlier above model projections.

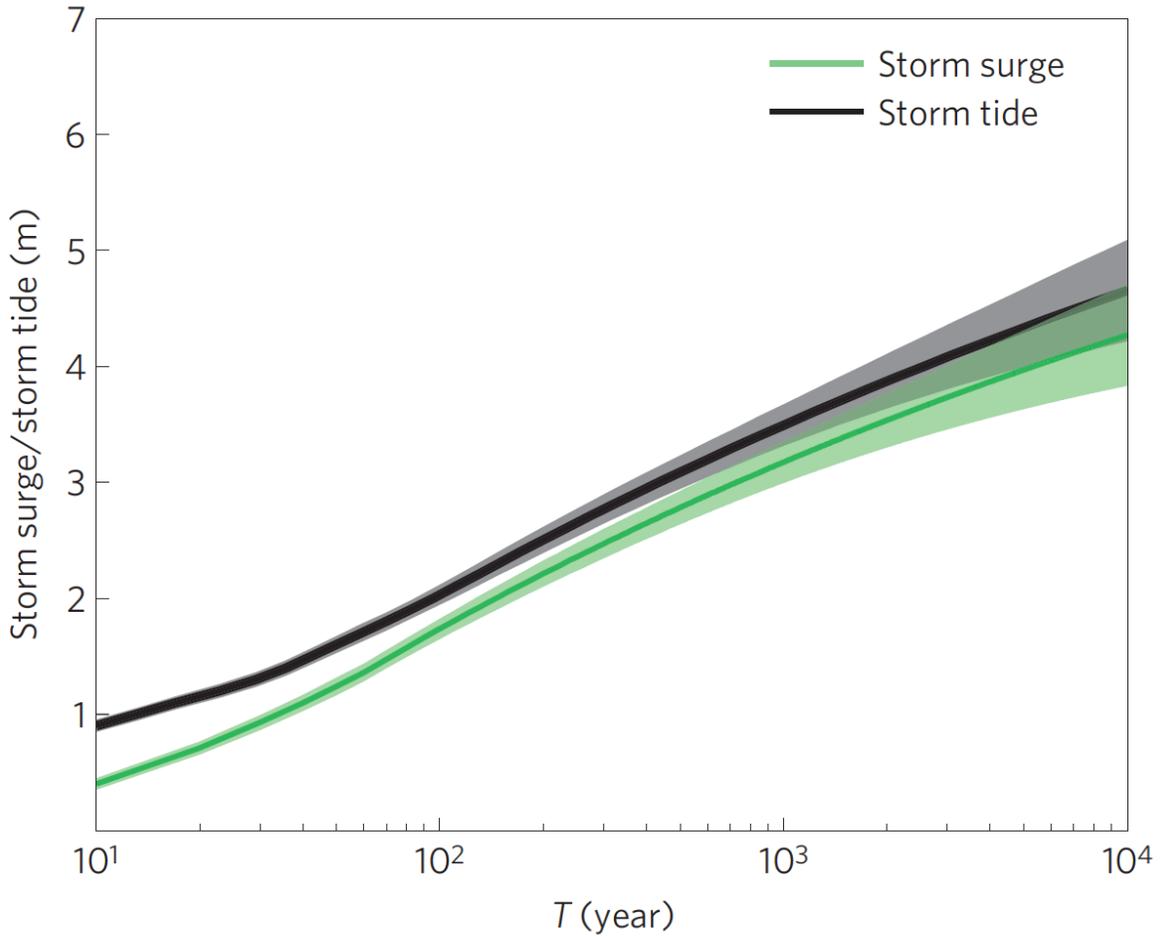

Figure SM2 Return level plot from (Lin et al., 2012) showing expected return levels for the Battery from synthetic storms used in (Aerts et al., 2013).

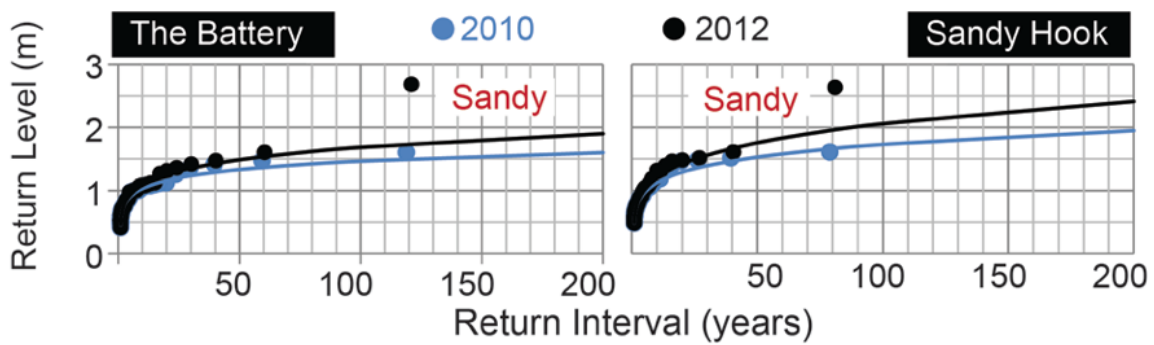

Figure SM3 Return level plot from (Sweet et al., 2013) showing estimated return levels (lines) for the Battery estimated by fitting the GEV distribution to observations (dots) for the tide gauge located at The Battery (black) and Sandy Hook (blue) tide.

GEV parameters used in this research were estimated in (Ceres et al., 2017) and produce an estimated return period for Surperstorm Sandy's storm surge of approximately 130 yrs. Fig. SM4. Illustrate the relatively good fit of annual block maximum tide gauge readings from The Battery Tide gauge and GEV model.

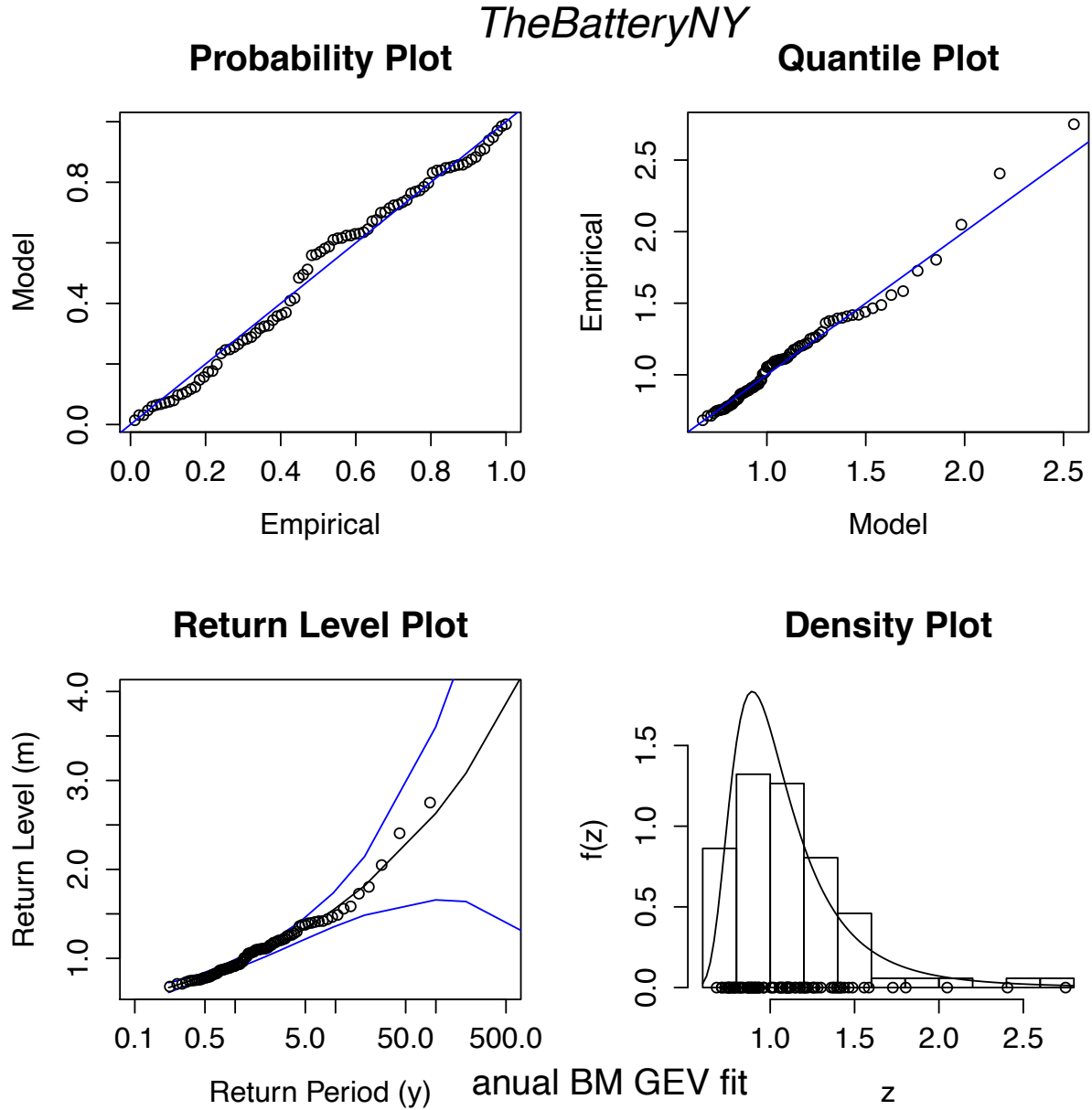

*Figure SM4 GEV equation goodness of fit and return level plots for tidal data from The Battery tide gauge. Panels a and b show probability and quantile comparison plots of GEV model and tide gauge data. Panel c shows modeled return level (black line), observations (black circles along the horizontal axis).*

Scileppi and Donahue's research examining sedimentary deposits near NYC (Scileppi & Donnelly, 2007) (see Fig. SM5.) suggest that other very large surges impacted NYC during the hurricanes of 1788, 1821, and 1893.

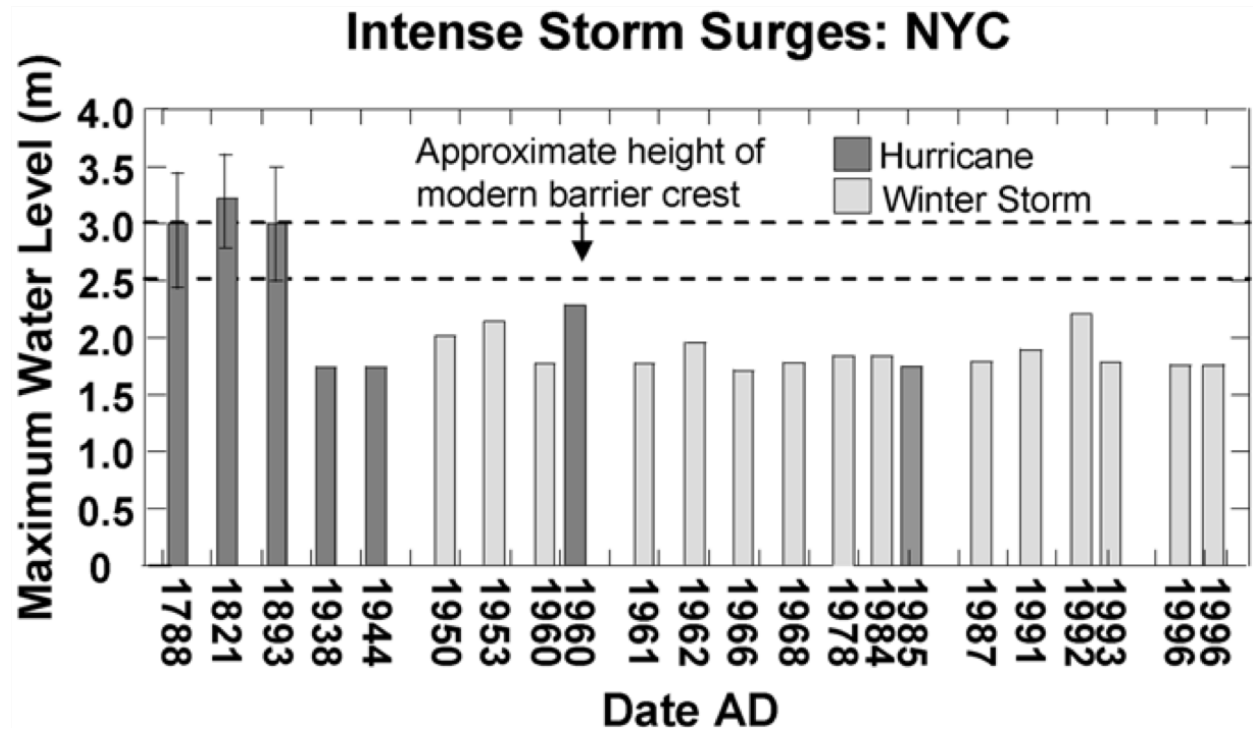

Figure SM5 Maximum water levels at The Battery inferred from sedimentary deposits from (Scileppi & Donnelly, 2007).

Observations from The Battery tide gauge during Superstorm Sandy, Hurricane Donna, and the `Great Appalachian' winter storm of 1950 illustrate the impact of the timing of the celestial high tide relative to the timing of the peak storm surge time. In the case of Surperstorm Sandy and Hurricane Donna the peak storm surge coincided with approximate high tide (see Figs. SM6 and SM7), whereas the `Great Appalachian' peak storm surge (see Fig. SM8 coincided with low tide. Variation in the timing and duration of storm surges relative to timing of the celestial tide introduces considerable uncertainty into projections of storm tides.

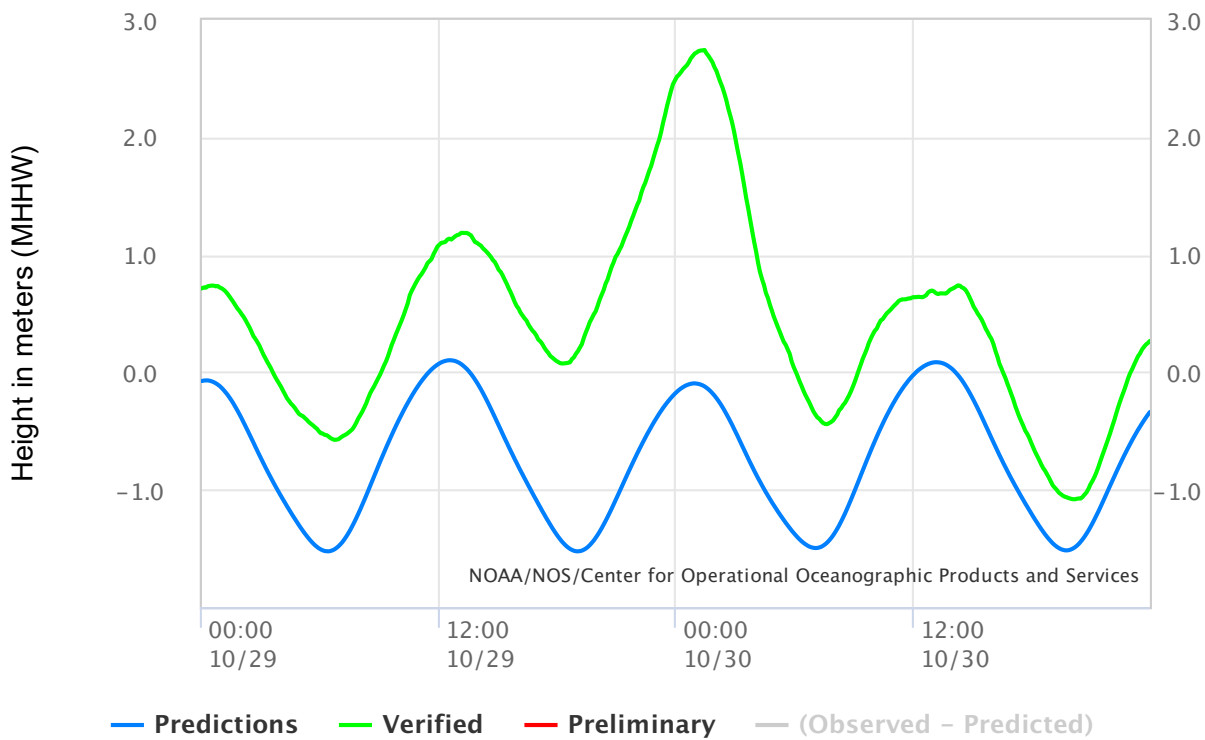

*Figure SM6 Tide gauge readings from The Battery tide gauge(*Water Levels - NOAA Tides & Currents*, n.d.) showing the observed tides relative to MHHW during Superstorm Sandy. Observed hourly tide gauge readings are in green, predicted celestial tides are shown in blue. The peak storm surge of 2.8 m occurred nearly simultaneously with celestial high tide, resulting in a peak storm tide of 2.7 m.*

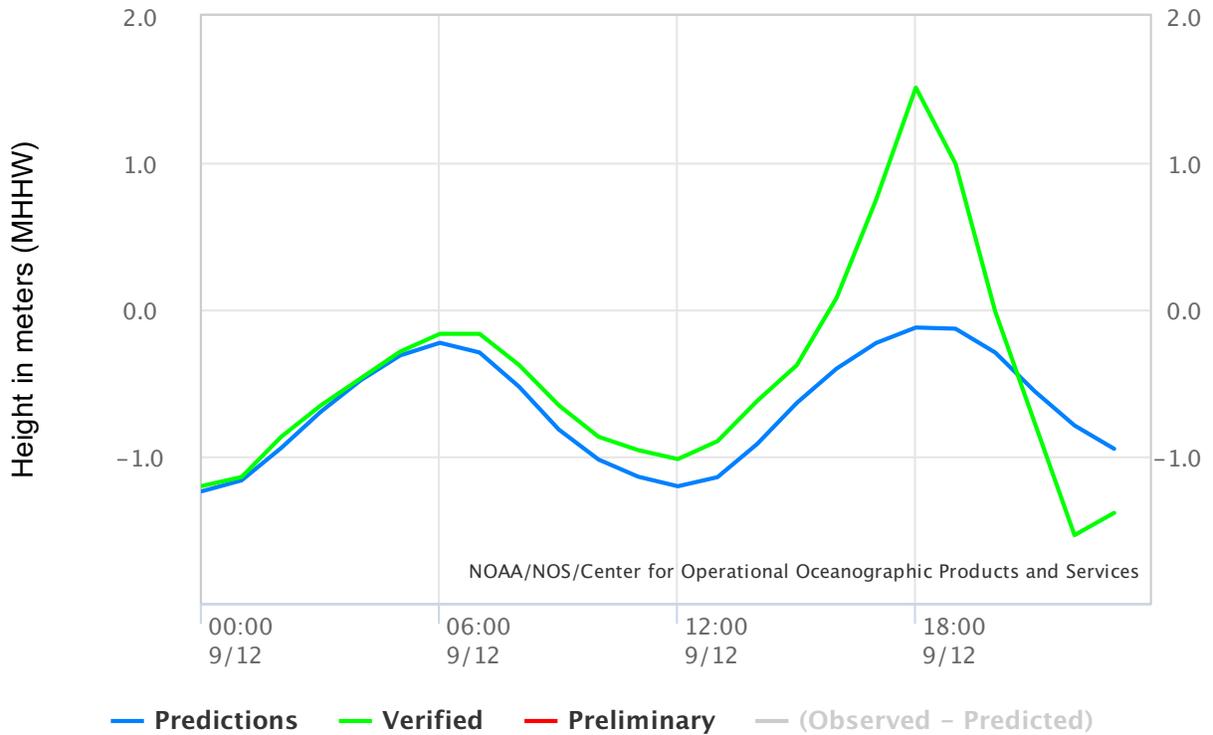

*Figure SM7 Tide gauge readings from The Battery tide gauge(*Water Levels - NOAA Tides & Currents*, n.d.) showing the observed tides relative to MHHW during hurricane Donna. Observed hourly tide gauge readings are in green, predicted celestial tides are shown in blue. The peak storm surge of 1.6 m occurred nearly simultaneously with celestial high tide, resulting in a peak storm tide of 1.5 m.*

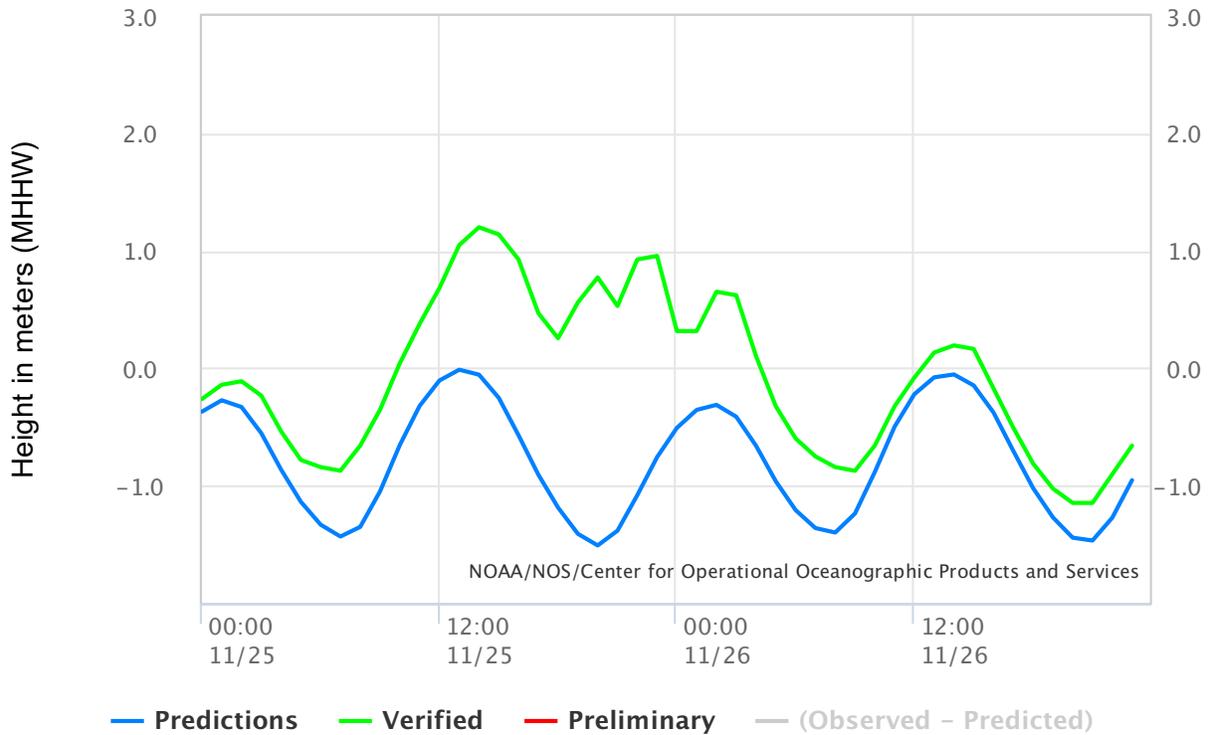

*Figure SM8 Tide gauge readings from The Battery tide gauge(NOAA, n.d.) showing the observed tides relative to MHHW during the `Great Appalachian' winter storm. Observed hourly tide gauge readings are in green, predicted celestial tides are shown in blue. The peak storm surge of 2.3 m occurred nearly simultaneously with celestial low tide. The peak storm tide of 1.2 m occurred near celestial high tide at the storm surge level of 1.2 m.*

Baseline GEV parameters used in this study from (Ceres et al., 2017) are based on measurements of the storm surge, the storm tide direct observation from The Battery tide gauge with the celestial tide and first order effect of sea-level rise removed. The largest storm tide and storm surge observed at the battery resulted from Superstorm Sandy. The second largest storm tide resulted from Hurricane Carol, and the third largest from the `Great Appalachian' winter storm. The rank of these storm surges, however, is reversed. The second largest surge resulted from the `Great Appalachian' winter storm, and Hurricane Carol's storm surge was the third largest surge. Using storm surge data rather than storm tide data (as in (Ceres et al., 2017) results in shorter return periods for very large storm surges.

The cumulative distribution function for the GEV(Coles, 2001) is,

$$F(x; \mu, \sigma, \xi) = exp\left[-\left(1 + \xi\frac{x-\mu}{\sigma}\right)^{\left(\frac{-1}{\xi}\right)}\right]. \tag{5}$$

We use GEV parameters estimated from The Battery tide gauge located adjacent to The Battery park in lower Manhattan(Ceres et al., 2017). The frequency, f(x) of an annual maximum surge, x is defined by three parameters: location $\mu$, scale ,$\sigma$ and shape, $\xi$. The location parameter ($\mu$) corresponds to the central, most likely portion of the distribution with respect to the block maxima storm surge heights. The scale parameter $\sigma$) describes the width of the distribution. The shape parameter ($\xi$) is associated with the skewness of the distribution. For this study, the location and scale parameters are non-stationary, while the shape parameter is stationary. The initial GEV parameters are $\mu_0$ = 0.936 m, $\sigma_0$ = $ 0.206 m, and $\xi_0$ =$ 0.232 m. Details on development of these parameters are explained in (Ceres et al., 2017). The location parameter increases by 0.05 m/year and the scale parameter increases by 0.000607 m/year, with each parameter contributing 0.5 meter per century linear increase to the 100-year surge return level. Given the other GEV parameters, this increase in the scale parameter corresponding to a 0.5 m per century increase in the 100-year storm surge.

Other studies have taken different approaches to determining the magnitude of storm surge risk to NYC. For example, (Sweet et al., 2013) fits storm tide observations to the GEV and finds a return period of 1,570 years for Superstorm Sandy's storm surge. Similarly, the return period for a Superstorm Sandy scale surge from the synthetic storms generated in (Lin et al., 2012) and subsequently used in (Aerts et al., 2013) (Aerts et al., 2014) is between 500 and 1,000 years. As can be seen in Figs. SM2 and SM3, the observation of Sandy's surge stands out (at least visually) as an outlier. Fig. SM4 shows the good fit between modeled GEV parameters from section SM1.2 used in this study and tide gauge observations at the battery(Ceres et al., 2017), especially in regards to the observation of Sandy's storm surge. Applying parameter values from (Ceres et al., 2017) here, Sandy's storm surge return period at the Battery tide gauge in lower Manhattan is 130 years. (Lin et al., 2012) and (Aerts et al., 2013, 2014) not provide GEV parameters for their estimated storm surges, but from examining supplemental Figs. SM2 and SM4, it is evident that return periods for storm surges used in article are always shorter than those used in (Aerts et al., 2013, 2014). The GEV shape parameter used in this article is positive, whereas the near linear relationship between return level and log return period in Fig. SM2 suggests that, were theses synthetic surges fitted to a GEV distribution, the shape parameter would be close to zero.

SM4 Data visualization tools

Once objectives are optimized and a population of Pareto dominant candidate cites is produced, decisionmakers face the difficult task of balancing the often competing objectives and priorities of different stakeholder groups. The iCOW framework can assist in this process to the extent that it both helps stakeholders identify strategy solutions that meet their preferred objectives and helps policy makers articulate and balance the preferences of their constituent stakeholder groups. The iCOW population of Pareto dominant candidate cities, however, is agnostic with respect to preference for different objectives. As a result, stakeholder groups may find the iCOW solution space is cluttered with candidate cities that they are not interested in, and this may detract from their ability to focus on the subsets of candidate cities with their preferred objective performance. Therefore, we discuss the importance of data visualization tools and techniques in section 3.1.

The J3 data visualization toolkit provides extensive controls for interacting with the iCOW framework objectives and the associated strategies and lever settings for the converged Pareto dominant solution cities.

SM4.1 Exploring the Manhattan iCOW objectives solution space

To illustrate this data visualization conundrum and to better understand the nature of this iCOW population converged Pareto dominant solution space, we examine the visible bifurcation of data that occurs at the lowest investment (and correspondingly highest damage) portions of Figs. 1, 2, and 3. the visible gap between the points closest to the Pareto front and the mass of points behind the front in Fig. 1. Other visually similar features can arise in the pareto optimal solution sets associated with optimization methods used and optimization settings, and policymakers and stakeholders might assume the three features are associated with similar underlying causes. Alternatively, if they cannot discern a reason for the presence of the visual feature, policymakers and stakeholders may assume that important features, such as the visual bifurcation of the data in Figs. 1-3 are insignificant artifacts of the data presentation or methods used to produce the data. The visual features are visible in the 3D representation of the data, but the projection provides little additional insight into the nature of the features. The features are not visible at all in the parallel axis plot, Fig. SM?? below.

Investigation into the data behind the bifurcated data, reveals that the split results from the tension between the benefit to cost ratio, total monetary net benefit, annual frequency of threshold events, and frequency of positive total monetary net benefit objectives. One family of solutions has low investment levels with high BCR. The framework achieves these objectives incorporating building resistance heights at lower (and therefore less expensive) resistance percentages. These investments always work to reduce damage, but the amount of damage reduced is capped by the volume of buildings made resistant. Moreover, the amount of reduced damage is relatively low, especially compared to the damage associated with the large surges of threshold events. As a result, these solutions have a relatively high annual frequency of threshold events. A second `family' of solutions are characterized by low threshold event frequencies, but lower BCR and higher investment levels required to achieve the same damage levels. These solutions are characterized by small dikes with relatively high base heights. These dikes act to reduce the damage of extreme storms below the threshold damage level, decreasing the frequency of these storms. At low investment levels, there are no Pareto optimal solutions between these families, resulting in the distinct visual bifurcation of the data visible in Figs. 1-3.

The gap between points at the cost vs. investment Pareto front and the mass of points behind the front are driven by a similar discontinuity in lever settings. Like the bifurcated case the feature is visible in both the 2D and 3D projections, Figs. 1 and 2, but not in the parallel axis plot, Fig. ??. In this case, the family of points on the front are characterized by dike only strategies with the dike located at the waterfront. The secondary front behind the gap is characterized by population members with combined strategies consisting of dikes offset from the seawall and the area between protected by different combinations of resistance height and resistance percent. As in the case of the bifurcated data, there are no Pareto optimal solutions between these two Pareto fronts.

Despite the `filaments' visual similarity to the bifurcated, and gaps in the data discussed above in both 2D and 3D plots, the `filaments,' are an artifact of the iCOW framework optimization and do not represent a characteristic of the underlying data describing the Pareto dominant solution population. Solutions along each `filament' have nearly equal benefit to cost ratio, whereas stepping from filament to filament results in a skip in BCR. Unlike the bifurcated data, there are Pareto optimal solutions between the `filaments,' the iCOW framework, however, did not identify those intermediate solutions due to the values of Borg MOEA parameters settings (discussed in more detail in the supplemental section 1.4 and table SM3. Changing these parameter values (specifically selecting a smaller epsilon value for the BCR) changes the spacing and location of the filament points but has a minimal effect on the location of the estimated Pareto front. We can think of the filaments as lines defining nearly equal BCR, and which, in the 3D plot, are locating the approximate position of the surface extending between the filaments.

SM4.2 Exploring Manhattan iCOW levers solution space

In some cases decisionmaker or stakeholder understanding of proposed solutions may be improved by viewing the converged Pareto dominant solution from the perspective of strategy levers. Additionally, stakeholders may be interested in other metrics, such as the cost associated with each lever setting, or the damage accrued by city zone, can be extracted from the solution space and displayed. Here we illustrate various projections that incorporate lever settings or information metrics.

SM4.3 Additional data visualization discussion and examples

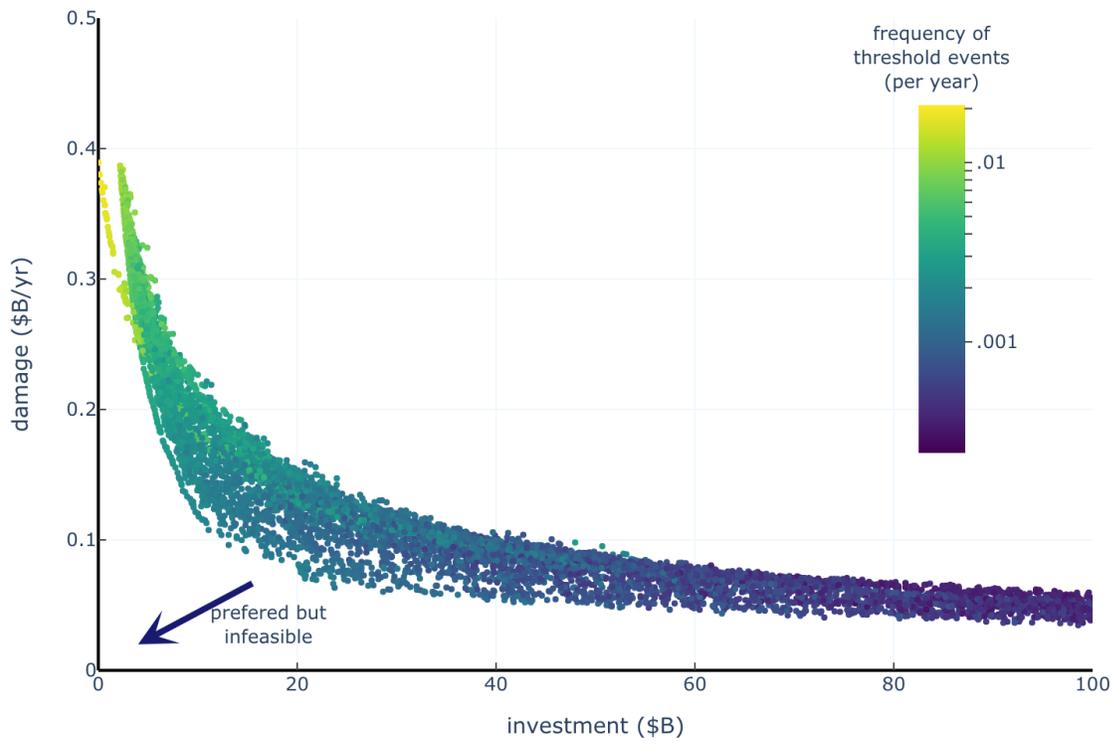

*Figure SM9 Two-dimensional data visualization of Manhattan iCOW emulation showing investment (x axis) vs. damage (y axis) for the Pareto dominant solution with investment cost less than $100 billion.*

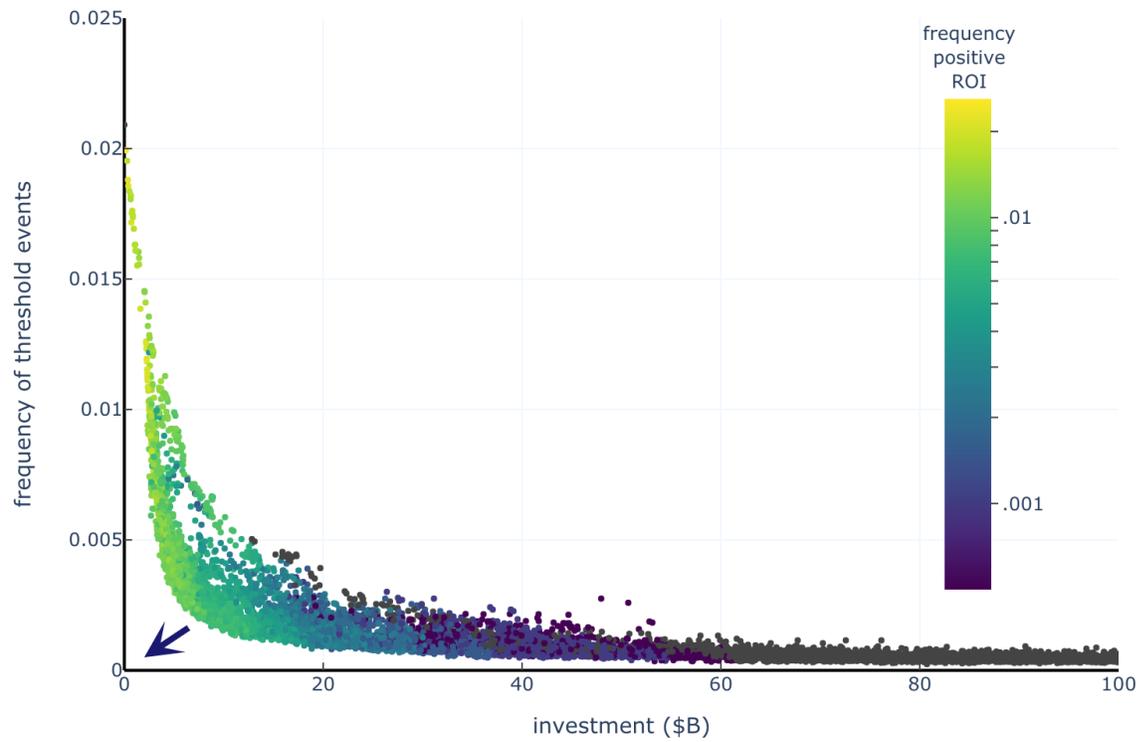

*Figure SM10 Two-dimensional data visualization of Manhattan iCOW emulation showing investment (x axis) vs. the annual frequency of threshold events) (y axis) for the Pareto dominant solution with investment cost less than $100 billion. The dark blue arrow indicates the location of the most preferred (but infeasible) solutions.*

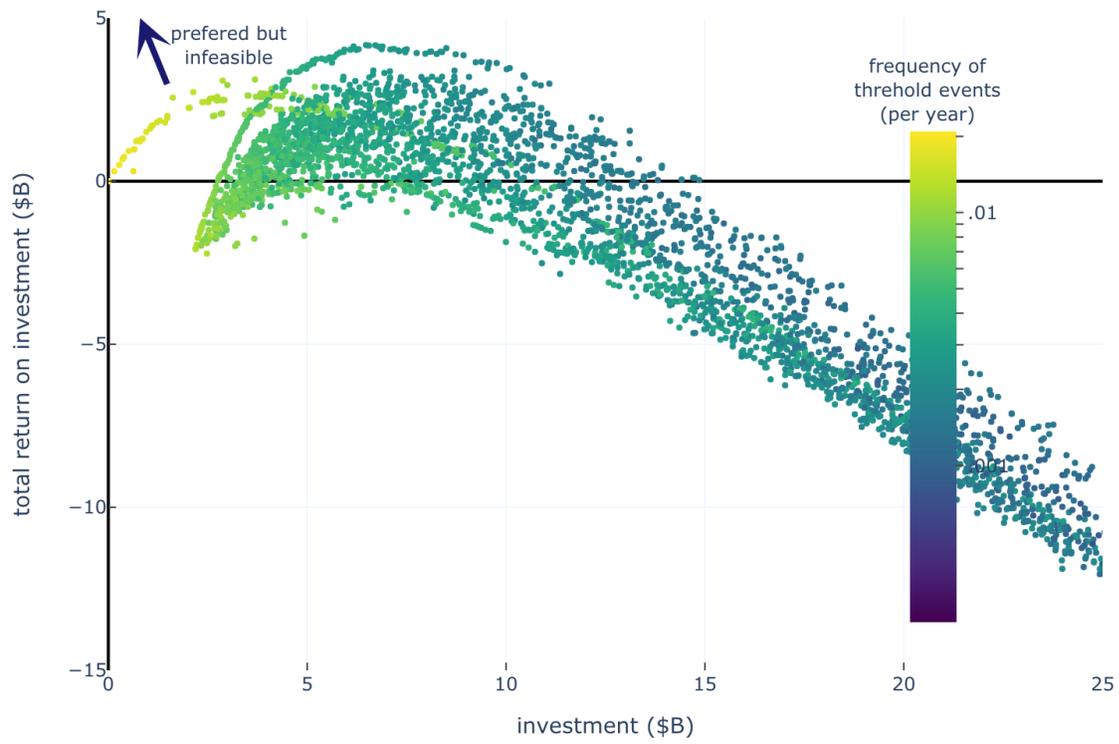

*Figure SM11 Two-dimensional data visualization of Manhattan iCOW emulation showing investment (x axis) vs. total return on investment (y axis) for the Pareto dominant solution with investment cost less than $25 billion.*

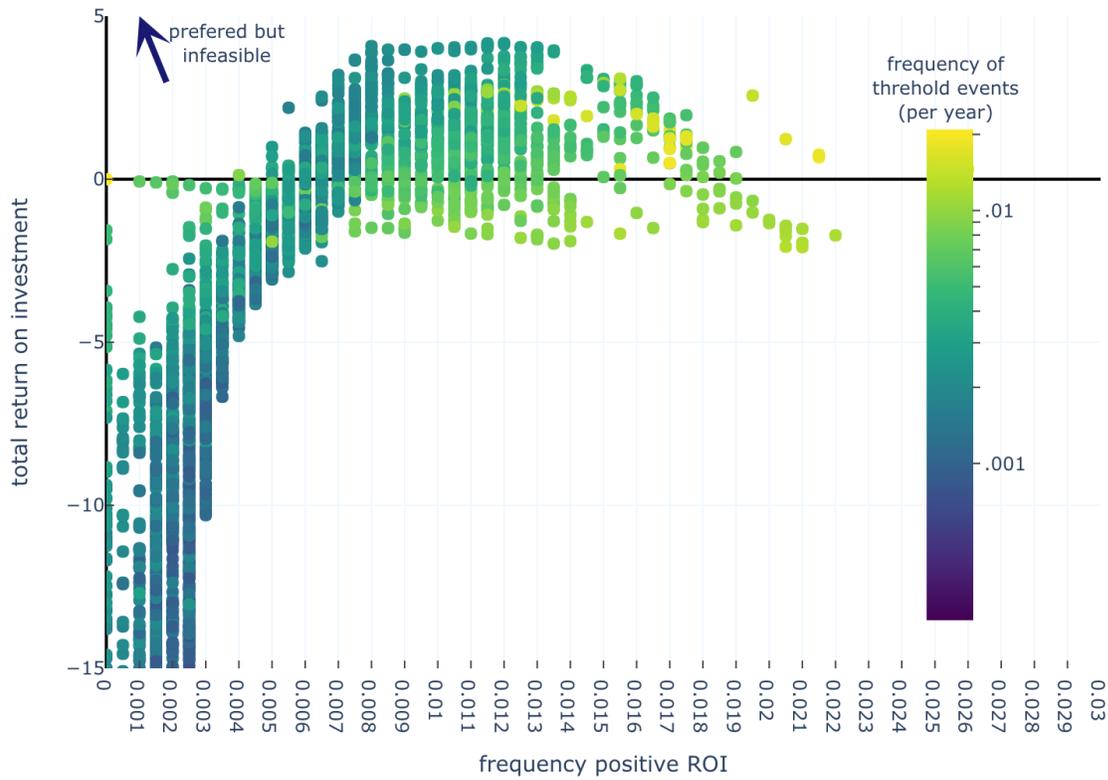

*Figure SM12 Two-dimensional data visualization of Manhattan iCOW emulation showing frequency of positive return on investment (ROI) (x axis) vs. total return on investment (y axis) for the Pareto dominant solution with investment cost less than $25 billion.*

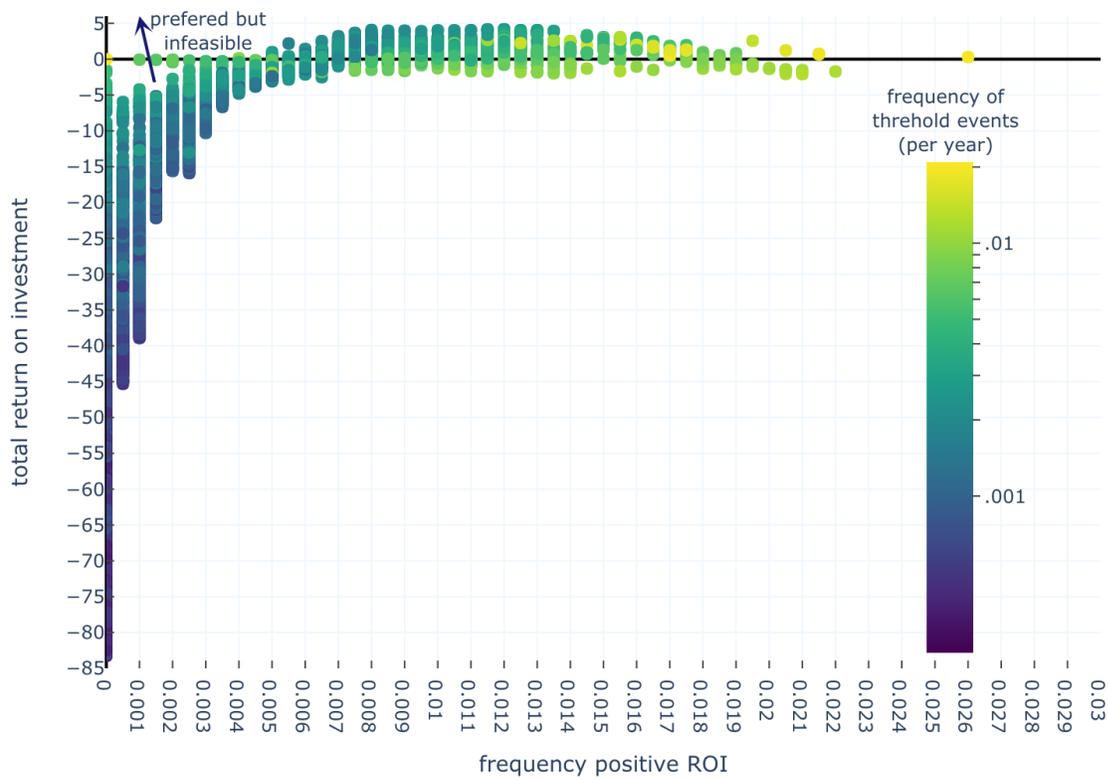

*Figure SM13 Two-dimensional data visualization of Manhattan iCOW emulation showing frequency of positive return on investment (ROI) (x axis) vs. total return on investment (y axis) for the Pareto dominant solution with investment cost less than $100 billion.*